\documentclass[%
  reprint,
 superscriptaddress,
 amsmath,amssymb,
 aps,
pra,
]{revtex4-2}
\usepackage{orcidlink}
\usepackage{graphicx}
\usepackage{physics}
\usepackage{xspace}
\newcommand{\ps}{phase space\xspace}
\usepackage{hyperref}
\usepackage{color}
\newcommand{\VEC}[1]{{\mbox{\boldmath${#1}$}}}

\usepackage[normalem]{ulem}

\makeatletter
\newsavebox{\@brx}
\newcommand{\llangle}[1][]{\savebox{\@brx}{\(\m@th{#1\langle}\)}%
  \mathopen{\copy\@brx\kern-0.5\wd\@brx\usebox{\@brx}}}
\newcommand{\rrangle}[1][]{\savebox{\@brx}{\(\m@th{#1\rangle}\)}%
  \mathclose{\copy\@brx\kern-0.5\wd\@brx\usebox{\@brx}}}
\makeatother

\usepackage{forloop,ifthen} 

\usepackage{xifthen}
\newcommand{\refAppendix}[6]{#1
  \ifthenelse{\isempty{#2}}%
    {}
    {\protect\cite{#2}}
    #3\protect\ref{#4}#5#6\xspace
}

\begin{document}

\title{Lindblad Superoperators from Wigner's Phase Space Continuity Equation}

\author{Ole Steuernagel\orcidlink{0000-0001-6089-7022}}
\email{Ole.Steuernagel@gmail.com}
\affiliation{Institute of Photonics Technologies, National Tsing Hua University, Hsinchu 30013, Taiwan}

\author{Ray-Kuang Lee\orcidlink{0000-0002-7171-7274}}
\email{rklee@ee.nthu.edu.tw}
\affiliation{Institute of Photonics Technologies, National Tsing Hua University, Hsinchu 30013, Taiwan}
\affiliation{Department of Physics, National Tsing Hua University, Hsinchu 30013, Taiwan}
\affiliation{Center for Quantum Science Technology, Hsinchu 30013, Taiwan}
 
\date{\today}
\begin{abstract}
  For a simple quantum system weakly interacting with the environment Wigner's 1932 formulation of
  quantum physics can be used to derive coupling to the environment using simple algebra. We show
  that the correct expressions, using coupling terms of `Lindblad form', are forced upon us. This is
  remarkable given that it took several decades before Lindblad's result was found in 1976.
\end{abstract}

\maketitle

\section{Introduction}

We prove that the time evolution of Wigner's quantum
distribution~$W$~\cite{Wigner_PR32,Hillery_PR84} obeys the continuity equation,
$\frac{\partial }{\partial {t}}W = - \VEC{\nabla} \cdot \VEC{J}$ with the Wigner
current~$\VEC{J}$~\cite{Bauke_2011arXiv1101.2683B,Ole_PRL13} and use this to derive the
Lindblad-form for a harmonic oscillator.

We approach coupling to the environment by na\"ively using the bosonic ladder operators
$\hat a^\dagger$ and $\hat a$ to describe quantum jumps between discrete levels and then combine the
resulting expressions into the form of a current~$\VEC{J}$.  The guidance provided by the continuity
equation enforces the correct form of terms describing weak coupling to the environment: they are of
Lindblad form.

In Sect.~\ref{sec:motivation} we review some history and further motivate this work. Then, in
Sect.~\ref{sec:Wign-Distr-Continuity}, we remind the reader of Wigner's formulation of quantum
theory and contrast it with the Schr\"odinger--von Neumann representation.  For hamiltonians that
can be Taylor-expanded, Sect.~\ref{sec:vec-J-from_Moyal} proves that in the Wigner formulation a \ps
current,~$\VEC{J}$, fulfilling a continuity equation, always exists. Using the jump operators
$\hat a^\dagger$ and $\hat a$ na\"ively, see Sect.~\ref{sec:Ja}, leads to an instructive failure. In
Sect.~\ref{secsec:_Lindblad}, we show how adhering to Wigner's continuity equation helps us to
construct the correct Lindblad superoperator.

\section{Motivation\label{sec:motivation}}

We learn in high school that natural phenomena are `quantum', they are discrete. When an electron
drops between discrete energy levels it creates a photon, a discrete package of light.
The electron has suffered a quantum jump and so has the light field.

Such processes were studied in considerable detail starting in the
1930s~\cite{Weisskopf_ZP30,Kikuchi__ZP30,Rzazewski_JPBAMOP92,Pfau_PRL94,Ole_PRA95,leong2016time}.

But oftentimes we want to ignore the wealth of detail provided by such investigations and focus on
fundamental (simple) structural  aspects. That is the approach we take here.

\subsection{From jumps to continuous evolution\label{subsec:JumpContinuous}}


With the inception of quantum mechanics a statistical theory of quantum dynamics was given by
Heisenberg and Schr\"odinger; in their representations quantum jumps do not feature explicitly, the
evolution of conservative systems is continuous.

But a system coupled to the environment can be excited by it and can lose excitations to it.  Here,
we exclusively discuss the special case that such interactions are described, respectively, by the
bosonic jump operators for system excitation,~$\hat a^\dagger$, or de-excitation,~$\hat a$.  Where
$\hat a^\dagger$ and~$\hat a$ are the harmonic oscillator ladder operators obeying the bosonic
commutation relation $[\hat a, \hat a^\dagger] = \hat 1$.
In quantum optics they are used to describe the addition,~$\hat a^\dagger$, or removal,~$\hat a$, of
a photon from a light mode~\cite{Titulaer_Glauber__PR66}.

The associated quantum jumps are observed when sufficiently discriminating measurements single out
final states which have sufficiently small overlap with initial states of the system
evolution~\cite{Nagourney_Dehmelt__PRL86}. And this can be described using measurement
theory~\cite{Cook_Kimble__PRL85}.

The measurement process, in particular, where to draw the boundary between system and measurement
apparatus, is, however, incompletely understood.

We therefore feel it is instructive to use the
perspective of Wigner's formulation of quantum dynamics in order to review the connection between
quantum jump operators, the coupling to the environment and the constraints imposed on such coupling
descriptions.

Na\"ively applying the operators~$\hat a^\dagger$ and~$\hat a$ does not work, see
Sect.~\ref{sec:Ja}. But combining pieces from~$\hat a^\dagger$ and~$\hat a$ to construct a Wigner
continuity equation guides us directly to the correct form, using Lindblad
superoperators~\cite{Braasch_PRA19}, see Sect.~\ref{secsec:_Lindblad}.

We feel that this is an astonishingly simple and transparent `derivation' of such an important
result that it is worth sharing. Lindblad derived it only in 1976~\cite{Lindblad_CMP76}, and in the
general case it is hardly straightforward to see~\cite{Lindblad_CMP76,Pearle_EJP12}. 
Yet, at least for the special case covered here~\cite{Braasch_PRA19}, Sect.~\ref{secsec:_Lindblad} shows the
Lindblad form is `easy to guess' when we are guided by Wigner's continuity equation.

\section{Wigner distribution and its Continuity Equation~\label{sec:Wign-Distr-Continuity}}

The conservative time-evolution of Wigner's quantum \ps distribution $W(x,p,{t})$, for a one-dimensional
continuous system, is governed by the \ps current,~$\VEC{J}$, and obeys a continuity
equation~\cite{Oliva_PhysA17}
\begin{eqnarray}\label{eq:continuity}
\frac{\partial W(x,p,{t})}{\partial {t}} + \VEC{\nabla} \cdot \VEC{J}(x,p,{t}) = 0 \; .
\end{eqnarray}
Here, $\VEC{\nabla} = (\partial/\partial_x, \partial/\partial_p)$ is the \ps divergence operator
with respect to position~$x$ and momentum~$p$, ${t}$ is time, and~$\VEC{J}=(J_x,J_p)$ has two
components and is a functional of $W$ and the system hamiltonian~$H(x,p)$.

\subsection{Why use the Wigner Formulation?~\label{sec:why-wign-distr}}

The conventional approach of using Schr\"odinger's or von~Neumann's equation forces us to
investigate complex-valued wave functions to represent the state of the system, this already poses a
stumbling block for the direct visualization of state changes.

Wigner's formulation of quantum theory, encapsulated by Eq.~(\ref{eq:continuity}), has the great
advantage of allowing us to visua\-lize quantum dynamics in \ps more directly since $W$ is
real-valued~\cite{Kakofengitis_EPJP17}.  Additionally, Wigner's formulation permits the study of its
real-valued \ps current, $\VEC{J}$~\cite{Bauke_2011arXiv1101.2683B}, and formal line integrals along
$\VEC{J}$ which yield momentary snapshots of `field lines', reminiscent of classical phase
portraits~\cite{Ole_PRL13}.


Wigner's continuity equation~(\ref{eq:continuity}) is the \ps equivalent of the conventional von
Neumann's equation~\cite{Oliva_PhysA17}, see Eq.~(\ref{eq:W_of_vNeumann}) below.

This gives us a differential operator as a function of~$x$,~$p$ and~$t$, to describe infinitesimal
changes in $W$, in the explicit and highly constraining form~$-\VEC{\nabla} \cdot \VEC{J}$.  It is
this form that made us choose the Wigner distribution and its continuity equation as the starting
point for our investigation.

\section{${\VEC J}(x,p,{t})$ from Moyal's bracket~\label{sec:vec-J-from_Moyal}}

We will now remind the reader of how Wigner's and Schr{\"o}dinger's representation of quantum theory
are connected mathematically~\cite{Zachos_book_21}.

Consider a single-mode operator,~$\hat O$, given in coordinate
representation~$\langle x-y| \hat O | x+y \rangle = O(x-y,x+y)$. To map to Wigner's \ps
formulation we employ the
Wigner-transform,~${\cal W}[\hat O]$~\cite{Hancock_EJP04,Cohen_LectureNotes18,Zachos_book_21},
\begin{align}\label{eq:WignerWeyl_Trafo}
  {\cal W}[\hat O](x,p) =
  \int_{-\infty}^\infty dy\; O(x-\frac{y}{2},x+\frac{y}{2})\; {\rm e}^{\frac{{\rm i}}{\hbar} p y}\, .
\end{align}

A settled terminology for these mappings does not exist in published
literature~\cite{Hillery_PR84,Hirshfeld_AJP02,Hancock_EJP04,Case_AJP08,Rasinariu_FP12,Cohen_LectureNotes18,Zachos_book_21}.
Here, we use the convention that the Wigner transform ${\cal W}$ maps an opera\-tor $\hat O$ to its
Wigner-symbol ${\cal W}[\hat O](x,p) = O(x,p) $~\cite{Zachos_book_21}, which is a function on
\ps. The inverse map, from a Wigner symbol to the associated normally-ordered~\cite{Hancock_EJP04}
operator $\hat O$ should be called its Weyl-symbol
${\cal W}^{-1}[O(x,p)] = \hat O$~\cite{Hillery_PR84}.

If $\hat O$ is a normalized single-mode density matrix~$\hat \rho$, then the associated
\emph{normalized} distribution in the Wigner repre\-sentation
is its Wigner symbol~$W(x,p) \equiv {\cal W}[\hat \rho]/(2 \pi \hbar)$.

Assuming that the hamiltonian ${\cal W}[ \hat H(\hat x, \hat p) ] = H(x,p)$ is smooth enough, namely,
has a global Taylor expansion, the Wigner transform of the von~Neumann time evolution equation
\begin{equation}\label{eq:W_of_vNeumann}
  {\cal W}\left[ \frac{\partial \hat \rho}{\partial {t}} = \frac{1}{{\rm i}\hbar} [\hat H, \hat
    \rho] \right]
  \end{equation}
is the Wigner-symbol
\begin{equation}\label{eq:moyal_motion}
  \frac{\partial W}{\partial {t}} = \{\!\!\{ {H} , W \}\!\!\}  = \frac{1}{\rm i \hbar} \left( H \star W - W \star H\right)\; ,
\end{equation}
where the Groenewold-$\star$-product~\cite{Groenewold_Phys46} is the \emph{Wigner symbol} of
operator concatenation ${\cal W}\left[ \hat H \circ \hat \rho \right] /(2 \pi \hbar ) = H \star W $.
It is given
by~\cite{Groenewold_Phys46,Zachos_book_21}
\begin{align}\label{Eq:GroenewoldStar}
  \star &\equiv \exp\left[\frac{i\hbar}{2} 
          \overleftrightarrow {\partial} \right]
          = \sum_{n=0}^{\infty} \frac{(i\hbar \overleftrightarrow {\partial})^n}{2^n n!}  
          \; ,
\end{align}
where we use the shorthand notations~$\frac{\partial}{\partial_z} = \partial_z$ and
$\overleftrightarrow{\partial} = \!\!\left( \overleftarrow{\partial_x} \overrightarrow{\partial_p} -
  \overleftarrow{\partial_p} \overrightarrow{\partial_x} \right)\!\!$,
with arrows indicating the `directions' of differentiation:
$f\overrightarrow{\frac{\partial}{\partial x}} g = g\overleftarrow{\frac{\partial}{\partial x}} f =
f \frac{\partial}{\partial x} g$.

In other words, the non-commutative nature of the composition of Hilbert space operators manifests
itself in the complicated and non-commutative structure of Groenewold's $\star$-product in the
Wigner formulation~\cite{Zachos_book_21}. The $\star$-product is the key-ingredient to map quantum
behaviour into \ps~\cite{Hancock_EJP04,Zachos_book_21}.

Hence, `Moyal's bracket',~$\{\!\!\{ {H} , W \}\!\!\}$~of
Eq.~(\ref{eq:moyal_motion}),~\cite{Moyal_MPCPS49,Zachos_book_21}
is of `Sine' form
\begin{align}\label{EqMoyalBraket}
  \{\!\!\{ f, g\}\!\!\} 
     = \frac{2}{\hbar} f(x,p) \sin\!\!\left[\! \frac{\hbar}{2} \!\!\left( 
        \overleftarrow{\frac{\partial}{\partial x}} \overrightarrow{\frac{\partial}{\partial p}}
         - \overleftarrow{\frac{\partial}{\partial p}} \overrightarrow{\frac{\partial}{\partial x}}
  \right)\!\!\right] g(x,p) \; .
\end{align}

A series expansion of Eq.~(\ref{eq:moyal_motion}), using expression~(\ref{EqMoyalBraket}), reveals that to
first order the Moyal bracket equals the Poisson bracket of classical mechanics:
\begin{align}\label{Eq:classicalLimit}
  \lim_{\hbar \downarrow 0} \{\!\!\{ {H} , W \}\!\!\} = H \overleftrightarrow{\partial} W = \{H , W \} \; .
\end{align}

\subsection{Proof that $\{\!\!\{ {H} , W \}\!\!\} = -{\VEC \nabla \cdot
    J}(x,p,{t})$ \label{subsec:proofMoyalDiv}}

If~$H(x,p)$ has a global Taylor expansion, evolution equation~(\ref{eq:moyal_motion}) can be
rewritten as the divergence of Wigner's \ps current,~$\VEC J$, of
Eq.~(\ref{eq:continuity})~\cite{Oliva_Kerr_18,Oliva_PhysA17}.

Let us prove this statement by induction:

The lowest order~(\ref{Eq:classicalLimit}) of Eq.~(\ref{eq:moyal_motion}) has the form
$ H \overleftrightarrow{\partial} W = (\partial_x H) ( \partial_p W) - (\partial_p H) (\partial_x W)
= \{H,W\} = \partial_p(W \partial_x H) - \partial_x(W \partial_p H) \equiv -{\VEC \nabla \cdot
  J_1}(x,p,{t}) $, yielding ${\VEC J_1 = (W \partial_p H,-W \partial_x H)^\intercal}$.

Analogously, we can decompose terms of  order $N+1$, in evolution equation~(\ref{eq:moyal_motion}),
in terms of terms of order $N$

$H \overleftrightarrow{\partial}^{N+1} W = (\partial_x H) \overleftrightarrow{\partial}^{N}
\partial_p W - (\partial_p H) \overleftrightarrow{\partial}^{N} \partial_x W = \partial_p
((\partial_x H) \overleftrightarrow{\partial}^{N} W) - \partial_x ( (\partial_p H)
\overleftrightarrow{\partial}^{N} W ) \equiv -{\VEC \nabla} \cdot {\VEC J_{N+1}} $, where
${\VEC J_{N+1} = ((\partial_p H) \overleftrightarrow{\partial}^{N} W, - (\partial_x H)
  \overleftrightarrow{\partial}^{N} W)^\intercal}$.

\noindent
This completes our proof.

Similarly, for multimode systems, tracing out degrees of freedom commutes with pulling out the
gradient opera\-tor in front. For example, consider a two-mode system with modes $a$ and $b$ and let
us focus on mode $a$, tracing out~$b$. This yields
${\rm Tr}_b \{H_{ab} \overleftrightarrow{\partial_{ab}} W_{ab}\} = {\rm Tr}_b\{ - \VEC \nabla_{ab}
\cdot \VEC j_{ab} \} = -\VEC \nabla_{a} \cdot \VEC j_{a}$, where
${\VEC j}_{a} = {\rm Tr}_b\{ \VEC \nabla_{b} \cdot \VEC j_{ab} \}$ has two components in mode
$a$, as desired. This result generalizes to terms of higher order
in~$\overleftrightarrow{\partial_{ab}}$ using the arguments presented in the previous paragraph,
for an application see Ref.~\cite{Ole_23_J_BeamSplitter}.

What about non-conservative dynamics, as encountered in dissipative systems? The multimode
considera\-tions just given, tracing out (the bath) modes, cover this case. For a weakly coupled
dissi\-pative system the Wigner current has been deri\-ved in analy\-tical form~\cite{Braasch_PRA19}
(and also been investigated experi\-men\-tally~\cite{Chen_PRA23}).

\subsection{Remark on the `--' sign in the Moyal bracket
  $\{\!\!\{ {H} , W \}\!\!\} = \frac{1}{\rm i \hbar} \left( H \star W \right.$ `--'
  $\left. W \star H\right)$~\label{subsec:MoyalMinus}}

It is the minus-sign in the Moyal bracket which is responsible for removing all terms from it that
do not carry derivatives of~$W$. This is, of course, inherited from the commutator in von~Neumann's
equation.  But there it might appear to be less fundamental since it might be understood to `just
arise' from the combination of phases of the unitary evolution operators for the bra- and ket-part
which form the commutator.

Viewing it from the Wigner \ps perspective, knowing that ${\partial_t} W $ has to equal terms of the
form~$-{\VEC \nabla} \cdot {\VEC J}$, reinforces that this is a fundamental feature.






\begin{figure}[b] \centering
  \includegraphics[width=4.19cm]{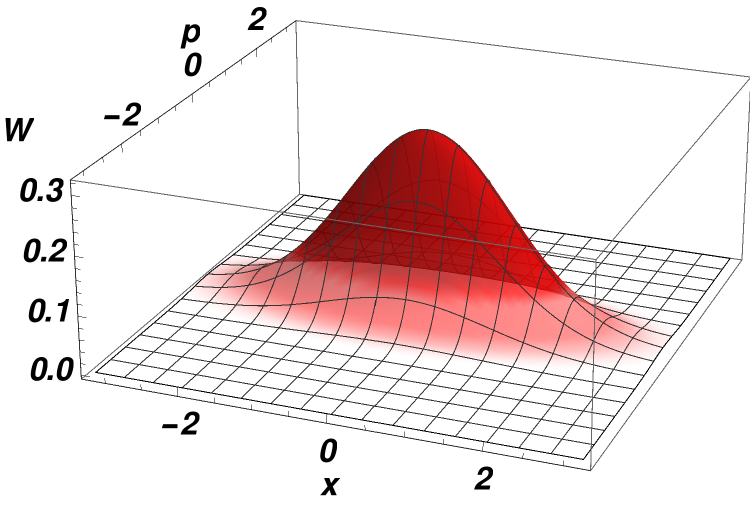}
  \includegraphics[width=4.19cm]{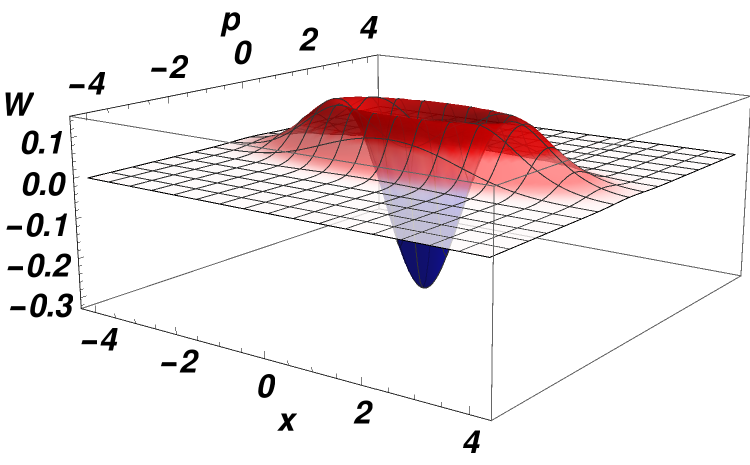}
  \caption{Wigner distributions~$W_V$ for weakly squeezed pure state (left) and the
    same state after formal photon addition,~$a^*  \star W_V \star a$, (right), the value
    of the normalized $x$-variance is $V=5/2$, for further details see Appendix~\ref{sec:Appendix_WDistributions}
    and  Ref.~\cite{Ole_24_AddingSubtractingSinglePhoton}.
    \label{fig:W_0_and_1}}
\end{figure}

\section{Photon addition: ${\cal W}[\hat a^\dag \hat \rho \; \hat a ]$ and
  $\VEC J_{\hat a^\dag \hat \rho \; \hat a}$ \label{sec:Ja}}

Let us now consider a bosonic single mode system, specifi\-cally a harmonic oscillator whose creation
operator is~$\hat a^\dag(\hat x, \hat p)$ $=$
$\sqrt{\frac{m \omega }{2 \hbar }} (\hat x-\frac{i \hat p}{m \omega })$. This can describe
an optical mode and its excitations~\cite{Titulaer_Glauber__PR66}. 

Setting $m$=1, $\omega$=1 and $\hbar$=1, the corresponding Wigner symbols are
${\cal W}[\hat a^\dag]=a^*(x,p)=(x-ip)/\sqrt{2}$ and ${\cal W}[\hat a]=a(x,p)=(x+ip)/\sqrt{2}$.  The
associated photon addition operator applied to a Wigner distribution (see
Fig.~\ref{fig:W_0_and_1}) therefore is
\begin{subequations}
    \label{eq:addPhotonWoverallEq}
\begin{align}
    \label{eq:addPhotonW_singularPart}
  \!\! 
  \frac{{\cal W}[\hat a^\dag \hat \rho_0 \; \hat a ]}{2 \pi \hbar } = \; & a^*  \star W_0 \star a =  \; \frac{1}{2} \left(p^2+x^2+1\right) W_0 + \\ \label{eq:addPhotonW}
& + \left[ -\frac{1}{2} {\VEC \nabla} \cdot \left(\begin{array}{c} x W_0
                                                       \\
                                                       p W_0
                                                     \end{array}\right)
   + \frac{\VEC \Delta W_0}{8} \right]    \; .
  \end{align}
  \end{subequations}

The differential operator in the square bracket~(\ref{eq:addPhotonW})
obviously has the desired form of $ \VEC{\nabla} \cdot {\VEC J} $ where
$ {\VEC J} = \frac{1}{8}([-4x + \partial_x]W_0 , [-4p + \partial_p]W_0 )^\intercal$
and ${\VEC \Delta W_0} = \VEC{\nabla} \cdot (\VEC{\nabla} W_0)$ the Laplacian
in \ps.

We now seek a current ${\VEC J}$ such that
$ \VEC{\nabla} \cdot {\VEC J} \approx - \Delta {\cal W}[\hat a^\dag \hat \rho_0 \; \hat a ] / \Delta
{t}$ can be integrated over one unit of time giving us
expression~(\ref{eq:addPhotonWoverallEq}):~$W_0 - \int_0^1 dt \VEC{\nabla} \cdot {\VEC J} = a^*
\star W_0 \star a$.

We do not have in mind to use a trivial, unphysical interpolation such as
$(1-t) W_0 + t \frac{{\cal W}[\hat a^\dag \hat \rho \; \hat a ]}{2 \pi \hbar} $, with $t=0,...,1$.

The term~(\ref{eq:addPhotonW_singularPart}), ${\frac{1}{2}(p^2+x^2+1)} W_0$, is troubling, it cannot be
rendered in the form $\VEC{\nabla} \cdot \VEC{J}(x,p)$.


If we attempt moving the polynomial terms of~(\ref{eq:addPhotonW_singularPart}) from the outside to
the inside of a derivative operator we end up increasing the order of the polynomials in $x$ and $p$
that are involved.

For instance $ p^2 W = \partial_x \; (p^2 x \; W) - p^2 x \; \partial_x W$. But the second term shows
an unwelcome increase to a higher order polynomial (here $p^2 x$). Its polynomial degree, if we
similarly try to move it inside a derivative, will increase yet further; ad infinitum.

\subsection{Compensating unphysical terms~\label{sec:enforce_Lindbald}}

Similarly to $a^* \star W \star a$, the \ps expression for photon removal from the light field~$W$
has the form
\begin{subequations}
    \label{eq:removePhotonWoverallEq}
  \begin{align}\label{eq:a_W_aD_singular}
   a \star W \star a^* = & \; \frac{1}{2} \left(p^2+x^2-1\right) W + \\
& + \left[ +\frac{1}{2} {\VEC \nabla} \cdot \left(\begin{array}{c} x W
                                                       \\
   \label{eq:a_W_aD_divergence}
                                                       p W
                                                     \end{array}\right)
   + \frac{\VEC \Delta W}{8} \right]    \; .
\end{align}
\end{subequations}

Combining Eqs.~(\ref{eq:addPhotonW_singularPart}) and (\ref{eq:a_W_aD_singular}), we find, neither
$a^* \star W \star a + a \star W \star a^*$ nor $a^* \star W \star a - a \star W \star a^*$ can be
given in the form of the divergence of a current. 

Given that $a^*(x,p)$ and $a(x,p)$ are complex functions (representing the fact that $\hat a^\dag$
and $\hat a$ are not hermitian) we cannot use them by themselves, let us therefore consider the
simplest extension possible, for finding a current. We try quadratic combinations of $a^*$ and
$a$ to compensate for the unphysical terms identified above:

We know that the \ps dynamics of quantum harmonic oscillators behaves
classically~\cite{Oliva_PhysA17}:
 \begin{align}\label{eq:J_HarmOSC}
   a^* \star a \star W - W \star a^* \star a  = {\VEC \nabla} \cdot \left(\begin{array}{c} p \; W
                                                       \\
                                                      - x \; W
                                                                          \end{array}\right)    \; .
\end{align}
Instead of subtracting such terms, adding them gives us
 \begin{align}\label{eq:_nW+Wn}
   \!\!\!\!   a^* \star a \star W + & W \star a^* \star a  = \\ \label{eq:_nW+Wn_b}
   & \; \left(p^2+x^2-1\right) W - \frac{\VEC \Delta W}{4}    \; ,
   \\  \label{eq:_nDW+WnD}
   \hspace{-0.7cm}\text{and}\qquad  
   a \star a^* \star W + & W \star a \star a^*  = \\ \label{eq:_nDW+WnD_b}
                                    & \; \left(p^2+x^2+1\right) W - \frac{\VEC \Delta W}{4}    \; .
\end{align}
We see that expressions~(\ref{eq:_nW+Wn_b}) and~(\ref{eq:_nDW+WnD_b}) can
compensate the terms~(\ref{eq:addPhotonW_singularPart}) and~(\ref{eq:a_W_aD_singular}).
\\
Namely, (only) the combinations
 \begin{align}\label{eq:aD_W_a__Lindblad1}
   2  & a^*  \star W \star a - (  a \star a^* \star W +  W \star a \star a^* )=
   \\   \label{eq:a_W_aD__Lindblad1b}
&  \left[ - {\VEC \nabla} \cdot \left(\begin{array}{c} x W
                                                       \\
                                                       p W
                                                     \end{array}\right)
   + \frac{\VEC \Delta W}{2} \right] 
   \; , \\
  \label{eq:a_W_aD__Lindblad2}
   \hspace{-0.4cm}\text{and}\qquad   2  & a \star W \star a^* - (  a^* \star a \star W +  W \star a^* \star a )=
   \\\label{eq:a_W_aD__Lindblad2b}   
&  \left[ {\VEC \nabla} \cdot \left(\begin{array}{c} x W
                                                       \\
                                                       p W
                                                     \end{array}\right)
   + \frac{\VEC \Delta W}{2} \right] 
   \; ,
\end{align}
can be rendered as~$-{\VEC \nabla} \cdot {\VEC J_\mp}$, respectively, where
 \begin{align}\label{eq:J_minusplus}
  {\VEC J_\mp} = - \left(\begin{array}{c} \mp x + \frac{1}{2} \partial_x \; W
                                        \vspace{0.1cm} \\
                \mp p + \frac{1}{2} \partial_p \; W
                                                                          \end{array}\right)    \; .
\end{align}
Reinstating the variables $m$, $\omega$ and $\hbar$ this becomes
 \begin{flalign}\label{eq:aD_W_a__Lindblad1units}
   2  & a^*  \star W \star a - (  a \star a^* \star W +  W \star a \star a^* )=
   \\   \label{eq:a_W_aD__Lindblad1bunits}
&  \left[ - {\VEC \nabla} \cdot \left(\begin{array}{c} x W
                                                       \\
                                                       p W
                                                     \end{array}\right)
   + \frac{\hbar}{2 m \omega} \partial_x^2 W + \frac{\hbar  m \omega}{2} \partial_p^2 W \right] , \\
  \label{eq:a_W_aD__Lindblad2units}
   \hspace{-0.4cm}\text{and}\quad   2  & a \star W \star a^* - (  a^* \star a \star W +  W \star a^* \star a )=
   \\\label{eq:a_W_aD__Lindblad2bunits}   
&  \left[ + {\VEC \nabla} \cdot \left(\begin{array}{c} x W
                                                       \\
                                                       p W
                                                     \end{array}\right)
   + \frac{\hbar}{2 m \omega} \partial_x^2 W + \frac{\hbar  m \omega}{2} \partial_p^2 W \right].
\end{flalign}

\section{Lindblad form \label{secsec:_Lindblad}}

The standard Lindblad master equation terms
\begin{eqnarray}
\frac{d \rho}{dt}&=&-\frac{i}{\hbar}[H,\rho]+\frac{\gamma}{2}(\bar{n}+1)\left(2 a\rho a^{\dagger}-a^{\dagger}a\rho-\rho a^{\dagger}a\right)\cr
&&+\frac{\gamma}{2}\bar{n}\left(2a^{\dagger}\rho a-a a^{\dagger}\rho-\rho a a^{\dagger}\right)
\label{mastereq}
\end{eqnarray}
are well known~\cite{Braasch_PRA19}.

Including natural units, the associated currents~$ {\VEC J}_{\text{env}}$, associated with the
$\gamma$-terms in~(\ref{mastereq}) coupling to the environment, have the concise
form~\cite{Braasch_PRA19}
\begin{eqnarray}
  \label{eq:Jenv}
\!\!\!\!
\!\!\!\!
  {\VEC J}_{\text{env}} &=& -\frac{\gamma}{2} W \left(\begin{array}{c} x
                                                       \\
                                                       p
                                                     \end{array}\right) -\frac{\gamma}{2} \frac{\hbar}{\omega_0} (\overline{n}+\frac{1}{2}) \left( \begin{array}{c}  \partial_x W
                                                                                                                                                     \\  \partial_p W
                                                                                                                                                   \end{array}\right), \\
                        &\equiv& {\VEC J}_{\text{damp}} + {\VEC J}_{\text{diff}} \; ,
\end{eqnarray}
where~${\VEC J}_{\text{damp}}$ is the classical damping current, whereas the diffusive
current,~${\VEC J}_{\text{diff}} (\overline{n}=0)$, enforces Heisenberg's uncertainty
principle~\cite{Chen_PRA23}.

As such, our result for~${\VEC J_\mp}$ of Eq.~(\ref{eq:J_minusplus}), or
Eqns.~(\ref{eq:aD_W_a__Lindblad1units})-(\ref{eq:a_W_aD__Lindblad2bunits}), are not new, expressions
of open systems dynamics in \ps have been derived by Wigner-transforming
expressions such as~(\ref{mastereq})~\cite{Cabrera_PRA15,Braasch_PRA19}.

What is new, however, is that our derivation shows that the continuity equation of \ps quantum
dynamics~(\ref{eq:continuity}) enforces this form, and that it does so in a transparent way.

\subsection{Lindblad form put into context\label{secsec:_LindbladLiterature}}

For the specific coupling for photon creation~(\ref{eq:addPhotonWoverallEq}) or
annihilation~(\ref{eq:removePhotonWoverallEq}), our analysis essentially gives a rederivation of the
well known Lindblad form~(\ref{eq:aD_W_a__Lindblad1}) and~(\ref{eq:a_W_aD__Lindblad2}) of a simple
quantum system weakly coupled to a bath.

We think it is astonishing that the insistence to express dynamical terms as the negative divergence
of Wigner's \ps current,~$-{\VEC \nabla} \cdot {\VEC J}$, should have such far reaching and at the
same time specific consequences.

This is an interesting observation since the standard consistency requirements imposed on the
coupling terms~\cite{Lindblad_CMP76,Pearle_EJP12}, when using the Schr\"odinger--von Neumann
representation, do not bear resemblance to the requirement to be of the form of the divergence of a
current.

Of course, our analysis was restricted to one case (photon addition to, or subtraction from, a light
field) and, unsurprisingly, it is neither more general nor does it allow us to check on more subtle
consistency requirements, such as those elucidate in Ref.~\cite{Salmilehto_PRA12}.  It also does not
obviously enforce the fluctuation-dissipation theorem, namely, that all terms have to appear
simultaneously, as in Eq.~(\ref{eq:Jenv}) (also see~\cite{Chen_PRA23}).

A good introductory text for the theory of quantum damping using terms of Lindblad-form is
Ref.~\cite{Milburn_Walls__AJP83}, a study giving visual insight is given in Ref.~\cite{Katz_NJP08},
also see the many references in~\cite{Braasch_PRA19}.  Recently, using the \ps approach on which
this work is based, studies have been performed for various systems,
see~\cite{Cabrera_PRA15,Braasch_PRA19}. Using \ps moreover lends itself to studying dissi\-pative
dynamics with the help of the geometrical analysis provided by Wigner's \ps current,~$ {\VEC J}$,
see~\cite{Braasch_PRA19,Ole_23_J_BeamSplitter}, including recent experimental
work~\cite{Chen_PRA23}.

\section{Conclusions and Outlook\label{sec:Conclusion}}

We have found that the explicit requirement that quantum dynamics (in \ps) has to fulfil continuity
equation~(\ref{eq:continuity}) enforces that photon addition to, or subtraction from, a system has
to obey a Lindblad form~\cite{Braasch_PRA19,Chen_PRA23}. For this special case, the derivation
consists of a few lines of transparent calculations of extreme simplicity.

Using the standard (von Neumann) approach yields  more subtle implicit requirements, namely, trace-,
hermiticity- and complete positivity-conservation~\cite{Pearle_EJP12}.

When contrasting the two approaches we wonder, in light of the simplicity of the \ps approach,
whether the Lindblad form could have been discovered earlier, if Wigner's \ps formulation would have
been more widely adopted sooner~\cite{Zachos_book_21}.


\section*{Acknowledgements}

O.~S. thanks Dr. Hsien-Yi Hsieh for fruitful feedback.  This work is partially supported by the
National Science and Technology Council of Taiwan (Nos 112-2123-M-007-001, 112-2119-M-008-007,
112-2119-M-007-006), Office of Naval Research Global, the International Technology Center
Indo-Pacific (ITC IPAC) and Army Research Office, under Contract No. FA5209-21-P-0158, and the
collaborative research program of the Institute for Cosmic Ray Research (ICRR) at the University of
Tokyo.

\section*{Disclosures} The authors declare no conflicts of interest.

\begin{appendix}

\section{Wigner distributions displayed in Fig.~\ref{fig:W_0_and_1}\label{sec:Appendix_WDistributions}}

In Fig.~\ref{fig:W_0_and_1} we display the Wigner distribution $W_V$ of a pure squeezed state in
dimensionless units $m$=1, $\omega$=1 and $\hbar$=1 in terms of  variances $V$ relative to the
variance of the vacuum (harmonic oscillator ground) state. In other words, $V=1$ is the vacuum state
whereas for values of $V>1$, the state
\begin{align}\label{eq:Appendix:WsqueezedPure}  
W_V(x,p) =   \frac{1}{\pi} \exp [ - \frac{x^2}{V} - p^2 V ]  \; ,
\end{align}
describes a pure squeezed state squeezed in the $p$-direction and anti-squeezed in the
$x$-direction, compare Fig.~\ref{fig:W_0_and_1}, left panel.  The associated form of the
photon-added expression~(\ref{eq:addPhotonWoverallEq}) for $W_V$ (also see
Ref.~\cite{Ole_24_AddingSubtractingSinglePhoton}), displayed in the right panel of
Fig.~\ref{fig:W_0_and_1}, is
\begin{align}\label{eq:Appendix:WsqueezedPurePLUSphoton}
  a^* \star  W_V(x,p) \star a  = W_V(x,p)
  \left(2 p^2 V-1 + \frac{2 x^2}{V}\right) \; .
\end{align}

\end{appendix}

%
%

\end{document}